# Supporting Aging Well through Accessible Digital Games: The Supplemental Role of AI in Game Design for Older Adults

Brandon Lyman, MS[1], Yichi Zhang, MS[1], Celia Pearce PhD[1], Miso Kim, PhD[1],

Casper Harteveld, PhD, MS[1], Leanne Chukoskie, PhD[1,2], Bob De Schutter, PhD, MFA[1,3]

College of Arts, Media and Design[1], Bouvé College of Health Sciences[2], Khoury College of Computer Science[3]

**Northeastern University**

Boston, MA, United States

{lyman.br, zhang.yichi6, c.pearce, m.kim, c.harteveld, l.chukoskie, b.deschutter}@northeastern.edu

Corresponding Author: Brandon Lyman (lyman.br@northeastern.edu)

## Abstract

As the population continues to age, and gaming continues to grow as a hobby for older people, heterogeneity among older adult gamers is increasing. We argue that traditional *game-based* accessibility features, such as simplified input schemes, redundant information channels, and increased legibility of digital user interfaces, are increasingly limited in the face of this heterogeneity. This is because such features affect all older adult players simultaneously and therefore are designed generically. We introduce artificial intelligence – although it has its own limitations and ethical concerns – as a method of creating *player-based* accessibility features, given the adaptive nature of the emerging technology. These accessibility features may help to address unique assemblage of accessibility needs an individual may accumulate through age. We adopt insights from gerontology, HCI, and disability studies into the digital game design discourse for older adults, and we contribute insight that can guide the integration of player-based accessibility features to supplement game-based counterparts. The accessibility of digital games for heterogenous older adult audience is paramount, as the medium offers short-term social, emotional, psychological, cognitive, and physical that support the long-term goal of aging well.

**Supporting Aging Well through Accessible Digital Games: The Supplemental Role of AI in Game Design for Older Adults**

More adults aged 50+ are playing games now than ever, and the average age of gamers is increasing (Entertainment Software Association [ESA], 2024). Older adults are an especially heterogeneous group in terms of health, ability, personality, and social participation (Dannefer, 1988; Grigsby 1996; Light et al. 1996;). It follows that growth in this population brings with it divergent accessibility needs that are difficult to address due to their bespoke nature.

Current approaches to digital games accessibility for older adults alone are not robust enough to handle this increased heterogeneity. Although they provide a baseline level of accessibility that is paramount for many older gamers to play games at all, they fall short in that they cannot possibly account for the more nuanced and individual-specific barriers to gameplay. As a result, the growing population of older adults may not be able to surpass such barriers and may miss out on many of the benefits associated with digital gaming, many of which support an individual's ability to age well. We argue that a robust accessibility approach would encompass both game-based, general accessibility enhancements coupled with enhancements that are tailored to the unique needs of an older adult individual playing the game.

Artificial Intelligence (AI) is an emerging technology that has shown promise in improving accessibility outside of digital games (Chemnad & Othman, 2024; Hatami & Chegini, 2024; Mitre & Zeneli, 2024; Sain, 2024). Recent research suggests that AI can personalize accessibility experiences (Huang et al., 2024; Li and Liao, 2025; Mitre & Zeneli, 2024; Sain, 2024), which positions AI as a potential supplement to traditional accessibility solutions in digital games. Although this approach has its own flaws, and brings with it ethical concerns, it could help to mitigate the accessibility issues at the level of the individual rather than the game.

This article synthesizes the implications of growing heterogeneity among older adult gamers, the existing game-based approaches to accessibility noting their merits and limitations, the benefits older adults can obtain from digital gameplay, and how those benefits are supportive of aging well. We contribute generative insight through examples from HCI that suggest a player-based, AI-driven approach to game accessibility may be achievable with existing applications. Finally, we discuss the broader implications of introducing AI to digital games and accessibility discourse for older adult gamers.

## Heterogeneity in Aging

Analysis of recent demographic trends show that the global population is aging and will continue for the foreseeable future. A 2024 report noted that the number of people aged 60 and over will roughly double by 2050, from 1 billion in 2020 to 2.1 billion in 2050 (World Health Organization [WHO], 2024). The same report mentions that the proportion of the world population aged 60 and over will nearly double, from 12% in 2015 to 22% in 2050 (WHO, 2024), so the demographic is likewise growing compared to other demographics.

Alongside these global aging trends, there is evidence that older people are playing more digital games than ever before. The ESA (2024) reported that nearly 30% of all digital game players in the U.S. are over the age of 50, up from 17% in 2004, and 9% in 1999. Atop the frequency charts of why older adults choose to play games, 83% of Boomers and members of the Silent Generation who play games indicated they do so to pass the time or relax, 59% indicated they do so to have fun, and 61% indicated they do so to keep their brain and mind sharp (ESA, 2024). AARP (2024) reports that 45% of adults aged 50 and over play games at least once a month based on a 2023 tech trends report produced by the organization. The report also notes that many Gen Xers are already over 50, and millennials are not far behind, thus they expect the growing trend of aging gamers to prevail (AARP 2024).

Serving this growing population is challenging due to inherent diversity amongst older adults. The concept of aged heterogeneity is described in the late 1980's, in a paper that criticized the discourse at the time for ignoring a "seemingly obvious" phenomenon that people become more individualized as they age (Dannefer, 1988, p. 373). Building upon this foundation, Grigsby (1996) presents a short review of three studies that they feel properly capture aged heterogeneity in the domains of political economy, health, and personality. Finally, Light et al. (1996) investigate the drivers of aged heterogeneity from the life course perspective, concluding that heterogeneity is derived from the interaction between personality and social environments, concluding that people become "more who they are" as they age (p.171).

Recognition of the growing older adult gamer population, coupled with the complexity introduced by aged heterogeneity, motivates our consideration of player-based accessibility solutions through AI to supplement existing game-based accessibility solutions described in prevailing digital games literature. We will explore these game-based solutions, their merits, and their limitations in the next section.

**Game-Based Accessibility Features**

We use the term *game-based* to describe accessibility features that are implemented into the digital implementation of the game itself. Accessibility-driven decisions that are implemented into the game influence all who play it, despite their age or unique performance abilities. The Game design literature for older adults is primarily focused on improving the efficacy of these features.

For instance, Whitlock et al. (2011) suggest designing the game to be controlled by single handed game controllers, which are more friendly towards age-related changes in physical ability. Additionally, they recommend implementing an adaptive difficulty scheme to help compensate for differences in player performance capabilities (Whitlock et al., 2011). Gerling et al. (2012) present an extended model of digital game design that encourages designers to consider age when designing their user interface and core mechanics. Vasconcelos and Silva (2012) propose ten "rules of thumb" for digital gaming platform design for older adults, which include ways of easing interaction, transportability to aid those with physical decline, and age-friendly interfaces (p. 8). Marston (2013) suggests that game interactions should be simple, controlled by voice or gestures, and "relate to real world or life experiences" when designing games for older adults (p. 114). Gamberini et al. (2014) provide a plethora of design implications for aging players of digital games, including utilizing redundant channels for transmission of crucial information, decreasing steps required to complete an action, and provision of formal training or manuals to learn about the game (p. 298–301).

These accessibility design insights are critical in providing a baseline level of accessibility to many people with varying levels of physical and cognitive ability. Many encourage minimizing unnecessary complexity, which in our opinion is crucial to effective game design regardless. Such game-based features are a sound approach given the limitations of designing an entertainment product, especially for a more homogenous demographic such as teens and young adults. As heterogeneity increases with age, however, older adults' accessibility needs to trend away from general and may require a more individualized approach to fully access digital games.

One way to conceptualize the importance of the accessibility features is through McLaughlin et al.'s (2012) notion of a cost-benefit analysis performed by older adults when deciding to play a game. In this model, the costs associated with playing digital games include lack of control, initial frustration, usability challenges, poor design for aging, and feelings of clumsiness, among others (p. 16). Many of these costs are directly associated with games accessibility and are reduced by the efforts of game designers to implement game-based accessibility features. Moreover, the model contends that older adults exhibit less motivation to play digital games if the costs outweigh the benefits. By considering additional, *player-based* accessibility features, we seek to further reduce the weight on the negative side of the scale such that benefits outweigh the costs.

**The Benefits of Digital Gaming and Aging Well**

On the other side of McLaughlin et al.'s (2012) cost-benefit scale are the benefits of digital gaming. Their paper identifies self-esteem, physical activity, social interaction, and positive emotions as benefits that are considered by older adults in this equation (McLaughlin et al., 2012). Many of these benefits are consistent with the themes and goals of *aging well,* a concept coined by Waddel et al. (2025) that evolved from a synthesis of research on the concept of *successful aging*.

*Successful aging* is a term first coined by Rowe and Kahn (1997) to describe maintaining desirable physical, cognitive, and social functioning throughout the aging process. The original model prioritizes avoidance of disease and disability, high cognitive and physical function, and engagement in with life as indicators of successful aging (Rowe & Kahn, 1997, p.434). Several newer models of successful aging are built upon Rowe and Kahn's foundational model, synthesized by Waddel et al. (2025), the result of which extends successful aging to include elements such as resilience, spirituality, and autonomy, among others. We refer the reader to Waddel et al.'s (2025) work for a more comprehensive understanding of their synthesis.

Waddel et al. (2025) highlights the flaws in the successful aging model, noting that the labeling of factors that are beyond the control of an individual (such as disease and disabilities) as success or failure is an unjust dichotomy (p.6). Moreover, they contend that such a conceptualization of aging fails to consider individual context, which resonates with our proposal of *player-based* accessibility features supplementing game-based counterparts (Waddel et al., 2025).

To highlight the connection between digital gameplay and aging well, we briefly expand upon the benefits identified by McLaughlin et al., offering insight from more recent literature. Research suggests that digital games can provide beneficial effects for players, including social benefits, such as relationship building (Kaufman, 2017; Mohsin et al., 2021; Reer & Quandt, 2020;); emotional benefits such as alleviation of depression (Kaufman 2017), reduction of loneliness or isolation (Kaufman, 2017), leisure (Boot et al., 2020, Mohsin et al., 2021), and eliciting appreciation (Reer & Quandt, 2020); psychological benefits such as needs satisfaction, stress reduction, and personal fulfillment (Reer & Quandt, 2020); cognitive benefits such as subjective short-term memory improvements (Mohsin et al., 2021), executive function, and processing speed improvements (Zhang & Kaufman, 2016); and physical benefits, such as rehabilitation (Mohsin et al. 2021; Zhang & Kaufman, 2016).

These benefits are supportive of the elements of aging well discussed earlier. Creating new relationships via digital game communities helps older adults to continue engagement in social relationships, and this social support could likewise foster resilience to adversity. The emotional and psychological benefits are likewise supportive of the resilience required in the aging well framework, as they work to mitigate the impact of negative emotional responses while creating positive responses such as enjoyment or appreciation. Cognitive and physical benefits of digital games support good cognitive and physical function. From this synthesis, we contend that access to digital games supports an older adults' ability to age well, and that accessibility barriers to digital games prevent older adults from obtaining such benefits. More robust accessibility practices could reduce the costs of playing a game.

**The Case for Player-based Accessibility Features**

Whereas *game-based* accessibility features describe those implemented in the game itself, *player-based* accessibility features instead improve accessibility in a personalized and adaptive manner. One common example of a *player-based* accessibility feature is dynamic difficulty adjustment. In such an approach, the skills of the individual player are communicated to the system, and the system responds by changing the content to better suit those skills (Zohaib, 2018). While DDA is not necessarily driven by AI, it is an example of the kinds of player-based systems we contend will benefit gamers as they age. Literature on AI-based accessibility specifically for older adults in digital games is scarce. Through review of tangential literature from HCI and AI studies, we see opportunities for digital game design discourse to adopt AI for creating more player-based accessibility features.

Before discussing these opportunities, however, we wish to draw attention to the risks, ethical considerations, and limitations of implementing AI in this space.  In terms of risks to the individual, there are three pertinent negative consequences that could be experienced directly by users of AI accessibility solutions: lack of transparency, the potential for bias and discrimination, and data privacy risks.

The risks involved with lack of transparency is the inability to build trust between AI experts and users (Siau & Wang, 2020). Furthermore, any incomplete understanding of an AI algorithm leaves room for biases as the natural complexity of the algorithm helps to obscure biases from view (Eitel-Porter, 2021, p.74).

Bias and discrimination in AI algorithms are a result of the data they are trained on (Siau & Wang, 2020), which could include reflections of current and historical social inequalities (Shukla, 2024, p.29-30). The data the algorithm is trained on could also contain gender bias or racial bias that informs the AI application (Siau & Wang, 2020). An example of bias being perpetuated in AI systems, and pertinent to the demographic we are serving, is ageism, which takes on several forms in the AI space including inclusion of bias in digital datasets and exclusion of older users from AI technology (Stypinska, 2023). Propagating these biases on a large scale could further harm marginalized populations (Shukla, 2024, p.29).

Data privacy is a risk in the use of AI because such algorithms need copious amounts of training data to be effective, which can often be comprised of sensitive personal data (Zhou et al., 2020). As with any activity that involves collecting personal data, AI applications introduce a risk of privacy breach in which unauthorized users of the data leverage the information for unregulated and nefarious means, like identity theft (Piñeiro-Martín et al., 2023). More specifically to AI, the ability of the algorithm to recognize patterns could lead to deduction of information it does not directly have access to, like sexual orientation (Stahl, 2021).

The discourse around ethical AI is vast, and a comprehensive understanding of the discussion is beyond the scope of this article. In short summary, ethical AI is achieved for the individual when: (1) The system is *transparent* and *explainable* (Shukla 2024), (2) its usage *does not discriminate* against users based on “race, religion, gender, sexual orientation, disability, ethnic, (sic) origin or any other personal condition" (Barredo Arrieta et al., 2020, p.36), and (3) keeps the vast amount of user data needed to operate the model *private* (Wei and Liu, 2024).

The drawbacks associated with the use of AI will need to be deeply considered before implementing any player-based accessibility features aimed at reducing accessibility concerns on a large scale. Meanwhile, it is generative to examine how existing AI solutions could be applied to accessibility in digital games. We will do so by describing specific accessibility issues and demonstrating how AI has helped address each issue in fields other than digital game design. While we acknowledge that many types of accessibility barriers exist, for scoping purposes we chose to focus on cognitive and physical accessibility barriers.

**Example 1: Visual and Audio Processing**

Decline in sensory functions like vision and hearing can negatively impact a gaming experience. Impaired vision could make it more difficult to read text, identify objects, and detect game prompts. This could impair performance, engagement, and overall experience with digital games (Gamberini et al., 2014; Gerling et al., 2012; Rienzo & Cubillos, 2020; Vasconcelos & Silva, 2012). Reduced contrast sensitivity, slower visual adaptation, and slower processing of sensory information make it challenging for older adults to distinguish details in complex graphics or fast-moving scenes, as well as to adapt to changes in light intensities or contrasts in games (Harada et al, 2013). Additionally, reduced visual depth perception can make it hard for older adults to accurately estimate distances, specifically affecting their ability to interact with three-dimensional elements in games (Harada et al., 2013).

Compounding the challenges associated with the changes in visual function, hearing difficulties may present challenges to older adults when following verbal instructions, recognize audio cues, or distinguish between background music and important game sound effects. This can hinder the ability to accurately respond to audio-based tasks, leading to missed opportunities or slow reactions during gameplay (Cota et al., 2015). Taken together, the lower fidelity of visual and auditory information poses additional strain on the cognitive functions discussed earlier, as older adults may need to work harder to interpret visual or auditory information to compensate for sensory limitations, increasing cognitive load (Speekenbrink & Shanks, 2013), and potentially increasing the risk of cognitive decline (Maharani et al., 2018).

To address these accessibility concerns, a common use of AI in areas other than digital games is to make accessible visual information through audio. In a systematic literature review, Chemnad and Othman (2024) note that the majority of AI and accessibility discourse is focused on assisting with visual impairments. Kumar et al. (2024) provide a collection of on-the-market technologies to assist people with visual disabilities. They include often used applications like virtual assistants (Google Assistant, Cortana, and Siri), but also more specialized technologies such as screen readers (TalkBack and VoiceOver), and a visual identification assistant called Lookout (Kumar et al., 2024). Abdolrahmani et al. (2018) investigated user perception of these technologies, concluding from inductive thematic analysis of semi-structured interview responses that the assistive applications "enable participants to perform independently without the need for sighted assistance or specialized hardware/software" (p.251). Another empirical study employing reflexive thematic analysis of semi-structured interview and observation data found that a combination of Siri and VoiceOver had a strong, positive influence on maintaining independence for those with visual impairments (Sayago & Ribera, 2020). These findings suggest that, since existing technologies appear to be effective at assisting visually impaired people with non-specific, daily tasks, similar technologies could be likewise effective in assisting those experience age-related decline in vision in a more specific setting such as digital gameplay.

Less prevalent are visual-based accessibility solutions for those with reduced hearing ability. Hatami and Chegini (2024) developed an AI-driven system that converts speech to images and emojis to assist hearing-impaired individuals with understanding spoken language. The ethical constraints imposed by the creators on model parameters, as well as the focus on providing responses as close to real-time as possible (Hatami & Chegini, 2024), bodes well for the future of adapting such systems to operate ethically in the often fast-paced digital worlds of modern games.

#### Example 2: Motor Skills

The decline in motor skills can have a wide impact on the operational aspects of gaming for older adults. Often this change presents an initial barrier to gameplay through the controller itself. For example, if the controller is very small, the buttons are not well spaced, or if the game requires buttons to be pressed simultaneously, the control may act as a barrier for older adult gamers (Gamberini et al., 2014). A decline in fine motor skills could limit an older gamer's ability to make precise movements such as aiming or performing combos (Gerling et al., 2012, Dick & Overton, 2010).

Additionally, a decline in hand-eye coordination associated with age makes it more difficult for older adult gamers to perform actions quickly and accurately in games, thereby limiting overall performance (Cota et al., 2015; Gerling et al., 2012). Such a barrier could present itself in areas including operational precision, simultaneous actions (e.g., pressing multiple buttons or using dual joysticks), mistakes and missed actions (Gamberini et al., 2014).

The specialized physical demands of modern digital games may prohibitively challenge the declining motor skills of some older adults. Madaan and Gupta (2020) describe the development of assistive technology that combines voice commands, gestures, and a face-controlled computer mouse. This technology uses natural language processing to allow the user to interact with web sites with their voice. Additionally, a bespoke convolutional neural network is used for detecting hand gestures, with up to 94% accuracy with 27 trained hand gestures (Madaan & Gupta, 2020). No AI or machine learning was required to create the face-controlled computer mouse (Madaan & Gupta, 2020). Technologies such as Madaan and Gupta's could offer an alternative manner of interacting with digital games rather than traditional input devices that require small and precise inputs, such as a keyboard and mouse.

#### Example 3: Working Memory

Working memory is defined as the "temporary storage and process information required for cognitive tasks such as comprehension, learning, and reasoning" (Engle & Kane, 2003). According to Baddeley's (2000) model, working memory consists of four components correspondingly functioning to control related attention, retain verbal information, hold visual and spatial details, and integrate information across systems and time. This model suggests that declined working memory can present challenges in games such difficulty focusing on and switching tasks, forgetting instructions or dialogue, getting lost in visual environments, and struggling to follow the storyline or combine clues, etc.

Games often require players to retain and process information to engage in play, and sometimes players need to interact with many pieces of information simultaneously. A decrease in working memory increases multi-tasking demand, which could induce feelings of frustration for older adult gamers (Cota et al., 2015; Gamberini et al., 2014; Gerling et al., 2012). Even when the player is not multitasking, working memory decline may also hinder an older adult gamers' adaptability, especially when games require frequent task switching or decision-making. This could lead to a decrease in game motivation and enjoyment (Gerling et al., 2012, Rienzo & Cubillos, 2020). Additionally, reduced ability to analyze complex information may require older adults to expend more effort in processing and retaining of information for multi-step tasks or complex narratives, leading to more difficulty in decision-making processes (Dick & Overton, 2010).

Digital games also can demand intense working memory loads for some older adults. In an early development that could support working memory abilities in the future, Makhataeva et al. (2023) describe the implementation and user testing of an AI system (coupled with augmented reality) to help participants remember physical locations of objects. The study involved participants taking a walking tour of a 3-story lab facility. They were tasked with memorizing the location of several objects in the 3D space. The AI system was used to identify objects and map them to a 2D representation of the 3D space. Cognitive demand and effort greatly decreased with the use of the system, whereas temporal demand and frustration also decreased with a lower effect size (Makhataeva et al., 2023).

These findings suggest that the ability of AI systems to scan and interpret 3D visual information, as well as simplify these representations, can reduce the amount of cognitive effort required to complete spatial tasks. Additionally, this system was created with the Unity game engine, a publicly available tool for game development. This indicates that the adoption of this technology into a digital game environment could be straightforward, as it was developed using the same tools as digital games.

**Personalization**

Key to the idea that AI could be used to create player-based accessibility features is the personalization capabilities exhibited by the technology. A survey in education found that AI technologies promote learning accessibility by creating personalized education paths for users with varied needs (Sain, 2024). Mitre and Zeneli (2024) drew similar insights from a literature review conducted on the topic of AI in higher education. Research on AI models such as ChatGPT has found that such models are capable of developing user personas autonomously for the purpose of adaptive user experience and user interfaces (Huang et al., 2024). This highlights the ability of AI to personalize a user experience based on data from the user. In a more direct application of AI, Li and Liao (2025) created a digital game about fishing that used AI to tailor the content of the game in accordance with the performance of the older adult player. The adaptive application was shown to improve satisfaction and willingness to play (Li & Liao, 2025). It follows from these studies that AI could theoretically tailor an accessibility approach to a player. The approach could address accessibility issues, such as those described in the above examples, to the extent that the individual needs them addressed. The system could accomplish this based on personas created from the player's anonymous data, in a way that addresses their unique accessibility needs.

## Aid and Threat to Self-esteem

As a final piece to the puzzle of player-based digital game accessibility features, we consider the threat-to-self-esteem model (Fisher et al., 1982). This consideration further complicates the goal of reducing accessibility barriers to digital games. The main considerations that apply here are: (1) the characteristics of the entity providing the aid, the aid itself, and the perception of the aid as threatening or helpful, (2) if the aid is perceived as threatening, older adults may have a negative reaction such as lowered self-concept and (3) a distinction between accepting an offer of aid versus requesting of aid is important in predicting the impact on self-esteem (Fisher et al., 1982). In criticizing the threat to self-esteem model, Newsom (1999) proposes a relationship between five variables (helping characteristics, relationship variables, caregiver variables, care recipient variables, and situational variables), interactions with the caregiver, and negative reactions to care. A comprehensive understanding of each of these variables is outside the scope of this article, however the key takeaway is that the inherent complexity of human relationships makes it difficult to predict whether a reaction to aid from another human will be positive or negative. This underscores the need for a personalized approach to offering help. Moreover, in an AI system an algorithm, rather than another person, would be offering this help. This is likely to reduce the influence of the caregiver, relationship, and situational variables described by Newsom.

AI seems to address this concern gracefully, due to an AI model's ability to act independently of manual input. Based on Huang et al. (2024) and Li and Liao's (2025) experimental results, AI-driven changes to the game's user experience could be executed without an explicit request from the user. The threat to self-esteem model predicts that this would lead to more positive reactions to assistance, as "aid that must be requested may elicit threat because it involves a public admission of inferiority, whereas an offer of aid does not" (Fisher et al., 1982, p.40). Framing the activities of an AI accessibility agent as offering assistance would therefore be less likely to negatively affect self-image. This is preferable to the alternative of navigating a menu to search for the accessibility features of need, which could be viewed as a request for aid - or worse, requiring the assistance of another person to adjust the accessibility features for them, which would almost certainly be viewed as a request for aid.

## Conclusion

From this analysis, we conclude that AI-driven applications have the potential to address accessibility concerns of older adults at an individual level, recognizing and respecting the heterogeneity of older adults and limiting negative reactions to aid predicted by the threat to self-esteem model. This would comprise what we consider to be a player-based approach to digital games accessibility for older adults, which we maintain should be implemented in addition to traditional, game-based accessibility approaches prevalent in the literature. Although AI is not a brand-new technology, its entry into the modern zeitgeist is recent. The ethical and cultural impacts have not been studied to a great enough extent for us to recommend it as a solution to accessibility problems. More research is required to understand what designing a fair, private, and transparent AI solution for this space would look like. At a practical level, however, it seems to us that AI uniquely fits the demands of player-based accessibility solutions.

Further research is required on modern older adults' perceptions of AI to understand whether such a technology would be successfully appropriated if developed. Additionally, research on safe and ethical AI implementation is required as an AI-powered accessibility, as we have discussed, introduces several risks to users. Finally, older adults' opinions on what an AI-powered accessibility solution for digital games would look like should be paramount, and we suggest that research on the design for such an application be grounded in participatory design methodologies.

**Funding**

This work was supported by the National Science Foundation (NSF) [grant number 2427714]. Any opinions, findings, or conclusions expressed in this material are those of the authors and do not reflect the views of the NSF.

**Conflict of Interest**

We have no conflict of interest to declare.

**Acknowledgements**

AI (ChatGPT by OpenAI) was used to draft high-level outlines that were then refined by the authors. Additionally, the technology was used to brainstorm titles. No rhetoric produced by generative AI is present in the final manuscript.

**References**

AARP. (2024). *Age-Friendly Game Development: A Primer for Game Designers and Developers*. AARP. https://employerportal.aarp.org/age-inclusive-workforce/age-friendly-technology/age-friendly-gaming-development

Abdolrahmani, A., Kuber, R., & Branham, S. M. (2018). "Siri Talks at You": An Empirical Investigation of Voice-Activated Personal Assistant (VAPA) Usage by Individuals Who Are Blind. *Proceedings of the 20th International ACM SIGACCESS Conference on Computers and Accessibility*, 249–258. https://doi.org/10.1145/3234695.3236344

Baddeley, A. (2000). The episodic buffer: A new component of working memory? Trends in Cognitive Sciences, 4(11), 417–423. https://doi.org/10.1016/S1364-6613(00)01538-2

Barredo Arrieta, A., Díaz-Rodríguez, N., Del Ser, J., Bennetot, A., Tabik, S., Barbado, A., Garcia, S., Gil-Lopez, S., Molina, D., Benjamins, R., Chatila, R., & Herrera, F. (2020). Explainable Artificial Intelligence (XAI): Concepts, taxonomies, opportunities and challenges toward responsible AI. *Information Fusion*, *58*, 82–115. https://doi.org/10.1016/j.inffus.2019.12.012

Boot, W. R., Andringa, R., Harrell, E. R., Dieciuc, M. A., & Roque, N. A. (2020). Older adults and video gaming for leisure: Lessons from the Center for Research and Education on Aging and Technology Enhancement (CREATE). Gerontechnology, 19(2), 138–146. https://doi.org/10.4017/gt.2020.19.2.006.00

Chemnad, K., & Othman, A. (2024). Digital accessibility in the era of artificial intelligence—Bibliometric analysis and systematic review. *Frontiers in Artificial Intelligence*, *7*. https://doi.org/10.3389/frai.2024.1349668

Cota, T. T., Ishitani, L., & Vieira, N. (2015). Mobile game design for the elderly: A study with focus on the motivation to play. *Computers in Human Behavior*, *51*, 96–105. https://doi.org/10.1016/j.chb.2015.04.026

Dannefer, D. (1988). What's in a name?: An account of the neglect of variability in the study of aging. In *Emergent Theories of Aging*. https://www.researchgate.net/publication/232504288_What's_in_a_name_An_account_of_the_neglect_of_variability_in_the_study_of_aging

Dick, A., & Overton, W. (2010). Executive Function: Description and Explanation. In *Self. Soc. Regul. Dev. Soc. Interact. Soc. Underst. Exec. Funct.* (Vol. 7, pp. 7–34).

Eitel-Porter, R. (2021). Beyond the promise: Implementing ethical AI. *AI and Ethics*, *1*(1), 73–80. https://doi.org/10.1007/s43681-020-00011-6

Engle, R. W., & Kane, M. J. (2003). Executive Attention, Working Memory Capacity, and a Two-Factor Theory of Cognitive Control. In *Psychology of Learning and Motivation* (Vol. 44, pp. 145–199). Academic Press. https://doi.org/10.1016/S0079-7421(03)44005-X

Entertainment Software Association. (2024). *Essential Facts About The U.S. Video Game Industry* (Essential Facts, p. 32). Entertainment Software Association.

https://www.theesa.com/wp-content/uploads/2024/05/Essential-Facts-2024-FINAL.pdf

Fisher, J., Nadler, A., & Whitcher-Alagna, S. (1982). Recipient reactions to aid. *Psychological Bulletin*, *91*, 27-54. https://doi.org/10.1037/0033-2909.91.1.27

Gamberini, L., Raya, M. L. A., Barresi, G., & Fabregat, M. (2014). Cognition, technology and games for the elderly: An introduction to ELDERGAMES Project. *Ingeniería Del Agua*, *18*(1), ix. https://doi.org/10.4995/ia.2014.3293

Gerling, K. M., Schulte, F. P., Smeddinck, J., & Masuch, M. (2012). Game Design for Older Adults: Effects of Age-Related Changes on Structural Elements of Digital Games. *Entertainment Computing - ICEC 2012*, 235-242. https://doi.org/10.1007/978-3-642-33542-6_20

Grigsby, J. S. (1996). The Meaning of Heterogeneity: An Introduction. *The Gerontologist*, *36*(2), 145-146. https://doi.org/10.1093/geront/36.2.145

Gupta, S., Fua, K., Pautler, D., & Farber, I. (2013). *Designing Serious Games for Elders*. https://doi.org/10.13140/RG.2.1.5166.9848

Harada, C. N., Natelson Love, M. C., & Triebel, K. L. (2013). Normal Cognitive Aging. *Clinics in Geriatric Medicine*, *29*(4), 737-752. https://doi.org/10.1016/j.cger.2013.07.002

Hatami, M., & Chegini, M. (2024). Enhancing Digital Content Accessibility for the Hearing Impaired through AI-Driven Visual Representations. *2024 10th International Conference*

*on Artificial Intelligence and Robotics (QICAR)*, 322–328. https://doi.org/10.1109/QICAR61538.2024.10496621

Huang, Y., Kanij, T., Madugalla, A., Mahajan, S., Arora, C., & Grundy, J. (2024). *Unlocking Adaptive User Experience with Generative AI* (No. arXiv:2404.05442). arXiv. https://doi.org/10.48550/arXiv.2404.05442

Kaufman, D. (2017). Socioemotional Benefits of Digital Games for Older Adults. *Human Aspects of IT for the Aged Population. Applications, Services and Contexts*, 242–253. https://doi.org/10.1007/978-3-319-58536-9_20

Kumar, V., Barik, S., Aggarwal, S., Kumar, D., & Raj, V. (2024). The use of artificial intelligence for persons with disability: A bright and promising future ahead. *Disability and Rehabilitation: Assistive Technology*, *19*(6), 2415–2417. https://doi.org/10.1080/17483107.2023.2288241

Li, H.-H., & Liao, Y.-H. (2025). Application and effectiveness of adaptive in elderly healthcare. *Psychogeriatrics*, *25*(1), 1–13. https://doi.org/10.1111/psyg.13214

Light, J. M., Grigsby, J. S., & Bligh, M. C. (1996). Aging and Heterogeneity: Genetics, Social Structure, and Personality. *The Gerontologist*, *36*(2), 165–173. https://doi.org/10.1093/geront/36.2.165

Madaan, H., & Gupta, S. (2021). AI Improving the Lives of Physically Disabled. *Proceedings of the 12th International Conference on Soft Computing and Pattern Recognition (SoCPaR 2020)*, 103–112. https://doi.org/10.1007/978-3-030-73689-7_11

Maharani, A., Dawes, P., Nazroo, J., Tampubolon, G., Pendleton, N., & Sense-Cog WP1 group. (2018). Visual and hearing impairments are associated with cognitive decline in older people. *Age and Ageing*, *47*(4), 575–581. https://doi.org/10.1093/ageing/afy061

Makhataeva, Z., Akhmetov, T., & Varol, H. A. (2023). Augmented-Reality-Based Human Memory Enhancement Using Artificial Intelligence. *IEEE Transactions on Human-Machine Systems*, *53*(6), 1048–1060. IEEE Transactions on Human-Machine Systems. https://doi.org/10.1109/THMS.2023.3307397

Marston, H. R. (2013). Design Recommendations for Digital Game Design within an Ageing Society. *Educational Gerontology*, *39*(2), 103–118. https://doi.org/10.1080/03601277.2012.689936

McLaughlin, A., Gandy, M., Allaire, J., & Whitlock, L. (2012). Putting Fun into Video Games for Older Adults. *Ergonomics in Design*, *20*(2), 13–22. https://doi.org/10.1177/1064804611435654

Mitre, X., & Zeneli, M. (2024). Using AI to Improve Accessibility and Inclusivity in Higher Education for Students with Disabilities. *2024 21st International Conference on Information Technology Based Higher Education and Training (ITHET)*, 1–8. https://doi.org/10.1109/ITHET61869.2024.10837607

Mohsin, N. F., Jali, S. K., Bandan, M. I., Jali, N., & Jupit, A. J. R. (2021). Older Adults and Digital Game Trends, Challenges and Benefits. *2021 IEEE Asia-Pacific Conference on Computer*

*Science and Data Engineering (CSDE)*, 1–6. https://doi.org/10.1109/CSDE53843.2021.9718445

Newsom, J. T. (1999). Another Side to Caregiving: Negative Reactions to Being Helped. *Current Directions in Psychological Science*, *8*(6), 183–187. https://doi.org/10.1111/1467-8721.00043

Piñeiro-Martín, A., García-Mateo, C., Docío-Fernández, L., & López-Pérez, M. del C. (2023). Ethical Challenges in the Development of Virtual Assistants Powered by Large Language Models. *Electronics*, *12*(14), 3170. https://doi.org/10.3390/electronics12143170

Reer, F., & Quandt, T. (2020). Digital Games and Well-Being: An Overview. In *Video Games and Well-being* (pp. 1–21). Palgrave Pivot, Cham. https://doi.org/10.1007/978-3-030-32770-5_1

Rienzo, A., & Cubillos, C. (2020). Playability and Player Experience in Digital Games for Elderly: A Systematic Literature Review. *Sensors*, *20*(14), 3958. https://doi.org/10.3390/s20143958

Rowe, J. W., & Kahn, R. L. (1997). Successful Aging. *The Gerontologist*, *37*(4), 433–440. https://doi.org/10.1093/geront/37.4.433

Sain, Z. H. (2024). *Exploring the Benefits of Artificial Intelligence in Enhancing Learning, Accessibility, and Teaching Efficiency*. https://doi.org/10.5281/ZENODO.13968719

Sayago, S., & Ribera, M. (2020). Apple Siri (input) + Voice Over (output) = a de facto marriage: An exploratory case study with blind people. *Proceedings of the 9th International Conference on Software Development and Technologies for Enhancing Accessibility and Fighting Info-Exclusion*, 6–10. https://doi.org/10.1145/3439231.3440603

Shukla, S. (2024). Principles Governing Ethical Development and Deployment of AI. *International Journal of Engineering, Business and Management*, *8*(2), 26–46. https://doi.org/10.22161/ijebm.8.2.5

Siau, K., & Wang, W. (2020). Artificial Intelligence (AI) Ethics: Ethics of AI and Ethical AI. *Journal of Database Management*, *31*(2), 74–87. https://doi.org/10.4018/JDM.2020040105

Speekenbrink, M., & Shanks, D. R. (2013). Decision making. In *The Oxford handbook of cognitive psychology* (pp. 682–703). Oxford University Press. https://doi.org/10.1093/oxfordhb/9780195376746.013.0043

Stahl, B. C. (2021). Ethical Issues of AI. In B. C. Stahl (Ed.), *Artificial Intelligence for a Better Future: An Ecosystem Perspective on the Ethics of AI and Emerging Digital Technologies* (pp. 35–53). Springer International Publishing. https://doi.org/10.1007/978-3-030-69978-9_4

Stypinska, J. (2023). AI ageism: A critical roadmap for studying age discrimination and exclusion in digitalized societies. *AI & SOCIETY*, *38*(2), 665–677. https://doi.org/10.1007/s00146-022-01553-5

Vasconcelos, A., & Silva, P. A. (2012). *Designing tablet-based games for seniors: The example of CogniPlay, a cognitive gaming platform*.

Waddell, C., Van Doorn, G., Power, G., & Statham, D. (2025). From Successful Ageing to Ageing Well: A Narrative Review. *The Gerontologist*, *65*(1), gnae109. https://doi.org/10.1093/geront/gnae109

Wei, W., & Liu, L. (2024). Trustworthy Distributed AI Systems: Robustness, Privacy, and Governance. *ACM Comput. Surv.* https://doi.org/10.1145/3645102

Whitlock, L. A., McLaughlin, A. C., & Allaire, J. C. (2011). Video Game Design for Older Adults: Usability Observations from an Intervention Study. *Proceedings of the Human Factors and Ergonomics Society Annual Meeting*, *55*(1), 187–191. https://doi.org/10.1177/1071181311551039

World Health Organization. (2024, October 1). *Ageing and health*. https://www.who.int/news-room/fact-sheets/detail/ageing-and-health

Zhang, F., & Kaufman, D. (2016). Physical and Cognitive Impacts of Digital Games on Older Adults: A Meta-Analytic Review. *Journal of Applied Gerontology*, *35*(11), 1189–1210. https://doi.org/10.1177/0733464814566678

Zhou, J., Chen, F., Berry, A., Reed, M., Zhang, S., & Savage, S. (2020). A Survey on Ethical Principles of AI and Implementations. *2020 IEEE Symposium Series on Computational Intelligence (SSCI)*, 3010–3017. https://doi.org/10.1109/SSCI47803.2020.9308437

Zohaib, M. (2018). Dynamic Difficulty Adjustment (DDA) in Computer Games: A Review. *Advances in Human-Computer Interaction*, *2018*(1), 5681652. https://doi.org/10.1155/2018/5681652

## Tables/Figures

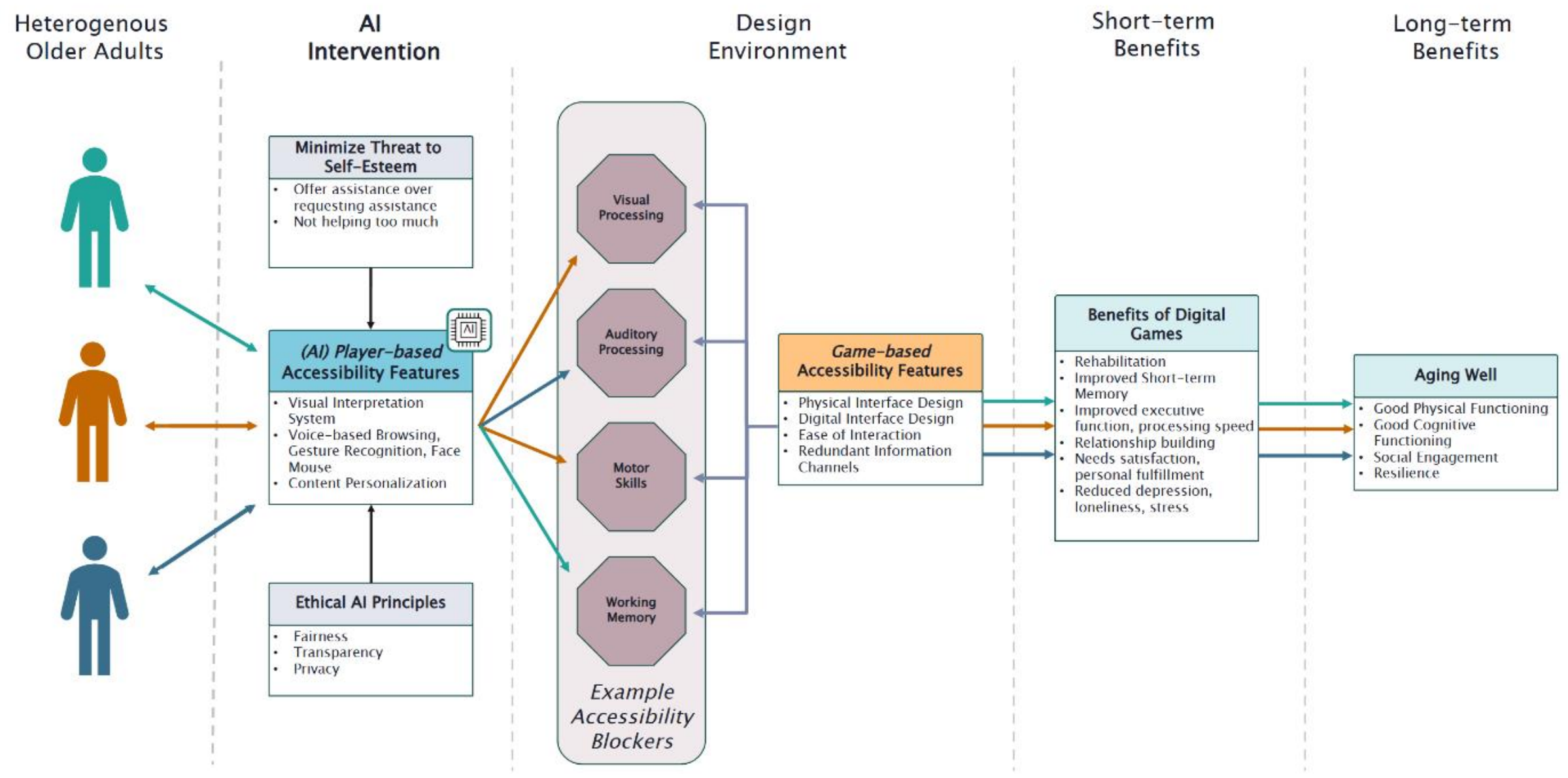

**Fig. 1.** This conceptual map demonstrates how AI-driven, *player*-based accessibility features are situated in relationship to older adult players. From left to right, heterogeneous older adults with unique accessibility needs (represented by different colors) interact with the *player-based* accessibility features. The work of Abdolrahmani et al. (2018), Hatami and Chegni (2024), Kumar et al., 2024, Madaan and Gupta, (2020), and Makhataeva et al. (2023) suggest such features are attainable with modern technology. Personalization capabilities of AI allow *player-based* features to operate within the unique context of the older adult and works to minimize threat to self-esteem (Fisher et al., 1982). Implementation of these features is guided by ethical AI principles. The *player-based* features and the *game-based* features work to reduce accessibility blockers from the player side and design side respectively. Lowering these barriers allows older adults to obtain the short-term benefits of digital games (represented again by colored arrows), supporting the longer-term goal of aging well.

**Alt text:** A conceptual map with three different colored human pictograms on the far left, with bilateral arrows indicating a relationship to player-based accessibility features, which is acted on by a block that says “Minimize Threat to Self-Esteem" on the top and “Ethical AI Principles” on the bottom. Moving from left to right, the arrows are pointing to stop-sign-like objects representing barriers to accessibility (visual processing, auditory processing, motor skills, and working memory). A box labeled “Game-based Accessibility features” has arrows pointing to each of the blockers from their right. Finally, colored arrows corresponding to the colors of the human pictograms point to a box titled “benefits of digital games,” which in turn has colored arrows pointing to a box titled “Aging well.”